# Invariants of Fokker-Planck equations

Sumiyoshi Abe[1,2,3]

[1]Physics Division, College of Information Science and Engineering,
Huaqiao Univeristy, Xiamen 361021, China

[2]Department of Physical Engineering, Mie University, Mie 514-8507, Japan

[3]Institute of Physics, Kazan Federal University, Kazan 420008, Russia

**Abstract.** A weak invariant of a stochastic system is defined in such a way that its expectation value with respect to the distribution function as a solution of the associated Fokker-Planck equation is constant in time. A general formula is given for time evolution of the fluctuations of the invariant. An application to the problem of share price in finance is illustrated. It is shown how this theory makes it possible to reduce the growth rate of the fluctuations.

In spite of a long history of the studies of stochastic processes and associated Fokker-Planck equations [1,2], the concept of invariants seems to have rarely been investigated, there. This fact may largely be due to the fact that such systems are dissipative and often have explicit time dependences in their microscopic dynamics, in general.

Recently, we have introduced and discussed the concepts of strong and weak invariants in time-dependent quantum systems [3]. A strong invariant is defined to be a Hermitian operator whose eigenvalues do not vary in time. A celebrated example of this type is the Lewis-Riesenfeld invariant of the time-dependent quantum harmonic oscillator [4]. On the other hand, a quantum weak invariant is a Hermitian operator whose eigenvalues vary in time but its expectation value remains constant. There, the dynamics is nonunitary in marked contrast to the case of the Lewis-Riesenfeld invariant, and a density matrix describing a quantum state obeys a dissipative master equation. The weak invariants are important, for example, in quantum thermodynamics due to the difference between isothermal and *isoenergetic* processes because of the quantum-mechanical violation of the law of equipartition of energy [5,6].

In this paper, we formulate the theory of weak invariants of classical stochastic systems. The expectation values of such quantities with respect to distribution functions as solutions of the associated Fokker-Planck equations remain constant in time, but their fluctuations do not. We present a general formula for the growth rate of the fluctuations. Then, we apply this theory to the problem of share price in finance as a simple example and show how it enables us to reduce the growth rate of the fluctuations of the invariant.

Let *X* be a random variable satisfying the following stochastic equation:

$$dX = f(X,t)dt + g(X,t)dW.  \tag{1}$$

Here, *f* and *g* are given functions of *X* and time *t*. $dW$ is assumed to be the Wiener process that satisfies Itô's rule [2]

$$(dW)^2 = dt. \tag{2}$$

Let *x* be a realization of the random variable, *X*. The probability of finding the value of *X* in the interval $[x, x+dx]$ at time *t* is denoted by $P(x,t)dx$, and the Fokker-Planck equation associated with the stochastic equation in Equation (1) is given by

$$\frac{\partial P(x,t)}{\partial t} = -\frac{\partial}{\partial x}\left\{\left[f(x,t) + \frac{\nu}{2}g(x,t)\frac{\partial g(x,t)}{\partial x}\right]P(x,t)\right\}$$
$$+ \frac{1}{2}\frac{\partial^2}{\partial x^2}\left[g^2(x,t)P(x,t)\right]. \tag{3}$$

Here, $\nu$ represents the calculus dependence in the multiplicative process, i.e., the *X* dependence of *g* on the right-hand side in Equation (1). For example, $\nu = 0\,(1)$ in the Itô (Stratonovich) calculus.

The weak invariant associated with the Fokker-Planck equation, $I(X,t)$, is a quantity, whose expectation value

$$\langle I \rangle = \int_{-\infty}^{\infty} dx\, I(x,t) P(x,t) \tag{4}$$

is constant in time, where $P(x,t)$ is a solution of Equation (3). Therefore, if $I(x,t)$ satisfies

$$\frac{\partial I(x,t)}{\partial t} + \left[ f(x,t) + \frac{v}{2} g(x,t) \frac{\partial g(x,t)}{\partial x} \right] \frac{\partial I(x,t)}{\partial x} + \frac{1}{2} g^2(x,t) \frac{\partial^2 I(x,t)}{\partial x^2} = 0, \qquad (5)$$

then it is a realization of the weak invariant.

It may be of interest to see how the variance of the weak invariant evolves in time. The time derivative of $(\Delta I)^2 = \langle I^2 \rangle - \langle I \rangle^2$ is calculated by the use of Equations (3) and (5) to obtain the following general formula:

$$\frac{d(\Delta I)^2}{dt} = \langle g^2 (I')^2 \rangle > 0, \qquad (6)$$

where $I' \equiv \partial I(X,t)/\partial X$. This shows how the variance monotonically increases.

Now, we apply the above discussion to the stochastic dynamics in finance [2]. Let $S(t)$ be the price of stock of a certain company at time $t$. The conventional stochastic equation for it reads

$$dS = \mu(t) S \, dt + \sigma(t) S \, dW, \qquad (7)$$

where $\mu$ and $\sigma$ are refereed to as the expected rate of return and volatility, respectively, and here both of them are allowed to depend on time. It is widely known that this equation offers a "dynamical" starting point in the Black-Scholes theory for option pricing (see Reference [2] for a nice explanation about this topic for physicists).

It is convenient to perform the change of the variable, $X = \ln S$, to transform the multiplicative process in Equation (7) to the following additive process:

$$dX = \omega(t)dt + \sigma(t)dW, \tag{8}$$

where

$$\omega(t) = \mu(t) - \frac{\sigma^2(t)}{2}. \tag{9}$$

Accordingly, Equation (5) becomes simplified to be

$$\frac{\partial I(x,t)}{\partial t} + \omega(t)\frac{\partial I(x,t)}{\partial x} + \frac{\sigma^2(t)}{2}\frac{\partial^2 I(x,t)}{\partial x^2} = 0. \tag{10}$$

Let us examine the following polynomial form for the invariant:

$$I(x,t) = \sum_{n=0}^{N} a_n(t) x^n. \tag{11}$$

Substituting this into Equation (10), we have

$$\dot{a}_N(t) = 0, \tag{12}$$

$$\dot{a}_{N-1}(t) + N\omega(t)a_N(t) = 0, \tag{13}$$

$$\dot{a}_n(t) + (n+1)\omega(t)a_{n+1}(t) + \frac{1}{2}(n+1)(n+2)\sigma^2(t)\, a_{n+2}(t) = 0$$

$$(n = 0, 1, 2, ..., N-2), \tag{14}$$

where the overdot implies the derivative with respect to time. The solutions of these equations can systematically be obtained for a given value of $N$. Clearly, $a_N$ should be a nonvanishing constant.

For example, we have in the case $N = 2$ that

$$a_2(t) = a_2(0), \tag{15}$$

$$a_1(t) = a_1(0) - 2 a_2(0) \Omega(t), \tag{16}$$

$$a_0(t) = a_0(0) - a_1(0) \Omega(t) + a_2(0) \left[ \Omega^2(t) - \Sigma^2(t) \right], \tag{17}$$

where $a_n(0)$'s are the initial values of $a_n(t)$'s, and

$$\Omega(t) = \int_0^t d\tau\, \omega(\tau), \qquad \Sigma^2(t) = \int_0^t d\tau\, \sigma^2(\tau). \tag{18}$$

Finally, let us evaluate the growth rate of the variance in Equation (6) in the case $N = 2$. The solution of Equation (3) with $f = \omega(t)$ and $g = \sigma(t)$ satisfying the initial condition, $P(x,0) = \delta(x - x_0)$, is given by

$$P(x,t) = \frac{1}{\sqrt{2\pi \Sigma^2(t)}} \exp\left\{ -\frac{1}{2\Sigma^2(t)} \left[ x - x_0 - \Omega(t) \right]^2 \right\}. \tag{19}$$

From this probability distribution function, the first and second moments are calculated to be: $\langle X \rangle = x_0 + \Omega(t)$, $\langle X^2 \rangle = \Sigma^2(t) + \left[ x_0 + \Omega(t) \right]^2$. The expectation value of the weak invariant, $I(X,t) = a_2(t) X^2 + a_1(t) X + a_0(t)$, is in fact constant:

$$\langle I \rangle = a_2(0)x_0^2 + a_1(0)x_0 + a_0(0), \tag{20}$$

where Equations (15)-(17) have been used. On the other hand, Equation (6) in the present case is found to be

$$\frac{d(\Delta I)^2}{dt} = 4a_2^2(0)\sigma^2(t)\Sigma^2(t) + \left[2a_2(0)x_0 + a_1(0)\right]^2 \sigma^2(t). \tag{21}$$

Therefore, if the initial values are chosen in such a way that

$$2a_2(0)x_0 + a_1(0) = 0 \tag{22}$$

holds, then the growth rate of the variance is reduced. This can be interpreted as an advantageous peculiar property of weak invariants in stochastic systems.

To summarize, we have discussed the concept of weak invariants of Fokker-Planck equations. Applying it to the problem of share price in finance, we have shown how the growth rate of fluctuations of the invariant can be reduced based on the structure of the invariant itself. Although we have considered here only a single random variable, generalization to multivariable cases is straightforward.

This work was supported in part by the High-End Foreign Expert Program of China, a Grant-in-Aid for Scientific Research from the Japan Society for the Promotion of Science (No. 26400391), and the Program of Competitive Growth of Kazan Federal University from the Ministry of Education and Science of the Russian Federation.

**References**

1. H. Risken, *The Fokker-Planck Equation*, 2nd ed. (Springer-Verlag, Berlin, 1989)

2. K. Jacobs, *Stochastic Processes for Physicists* (Cambridge University Press, Cambridge, 2010)

3. S. Abe, Phys. Rev. A **94**, 032116 (2016)

4. H.R. Lewis, Jr., W.B. Riesenfeld, J. Math. Phys. **10**, 1458 (1969)

5. C. Ou, S. Abe, EPL **113**, 40009 (2016)

6. C. Ou, R. V. Chamberlin, S. Abe, Physica A **466**, 450 (2017)